\documentclass[pra,twocolumn,showpacs,amsmath,amssymb,superscriptaddress,floatfix]{revtex4}
\usepackage{amssymb,amsmath}
\usepackage{graphicx}
\usepackage{dcolumn}
\usepackage{bm}
\def \PP {\mathcal{P}}

\def \CCC {\mathbb{C}}

\def \RRR {\mathbb{R}}
\def \tr {\mathrm{Tr}}
\def \hx {\hat{x}}

\def \hz {\hat{z}}

\def \equals {\ = \ }

\newtheorem{theorem}{Theorem}

\begin{document}

\title{Quantum noise thermometry for bosonic Josephson junctions in the mean field regime}

\author{Alex D. Gottlieb}
\affiliation{Wolfgang Pauli Institute, Nordbergstrasse 15, 1090 Vienna, Austria}
\author{Thorsten Schumm}
\affiliation{Wolfgang Pauli Institute, Nordbergstrasse 15, 1090 Vienna, Austria}
\affiliation{Atominstitut der \"Osterreichischen Universit\"aten, TU-Wien, Stadionallee 2, 1020 Vienna, Austria}
\date{\today}

\begin{abstract} 

Bosonic Josephson junctions can be realized by confining ultracold gases of bosons in multi-well traps, and studied theoretically with the $M$-site Bose-Hubbard model.    We show that canonical equilibrium states of the $M$-site Bose-Hubbard model may be approximated by mixtures of coherent states, provided the number of atoms is large and the total energy is comparable to $k_BT$.   Using this approximation, we study thermal fluctuations in bosonic Josephson junctions in the mean field regime.  Statistical estimates of the fluctuations of relative phase and number, obtained by averaging over many replicates of an experiment, can be used to estimate the temperature and the tunneling parameter, or to test whether the experimental procedure is effectively sampling from a canonical thermal equilibrium ensemble.

\end{abstract}
\maketitle
\section{Introduction}

Quantum degenerate Bose gases in double-well potentials exhibit coherent macroscopic tunneling dynamics, analogous to those in superconducting Josephson junctions \cite{SmerziEtAl, GatiEtAl-BJJ, LevyEtAl-BJJ}.  The observer can detect individual well populations by direct optical absorption imaging, and can infer the relative phase of the wave packets from
interference experiments \cite{EsteveEtAl-squeezing, ShinEtAl-DblW, SchummEtAlNaturePhysics}. Furthermore, atomic interactions can be tuned over a wide range by adjusting particle number and double-well parameters or by means of Feshbach resonances (cf. \cite{FeshbachReview} for a review).

Double-well systems are often modeled within a two-mode approximation by the Bose-Hubbard Hamiltonian.    
The parameters of the model are the number $N \gg 1$ of atoms,  the interaction energy $E_C$ for a pair of particles in the same potential well, and the tunneling coupling energy $E_J$ (cf. \cite{AnglinDrummondSmerzi, PitaevskiiStringari, GatiEtAl-NoiseThermometry}, which use the same notation).  
One distinguishes the ``Rabi", ``Josephson", and ``Fock" regimes \cite{Leggett} according to the relations 
\begin{eqnarray*}
E_C/E_J \  \ll \ N^{-2} && \hbox{(Rabi)} \\
N^{-2} \ \ll \  E_C/E_J \  \ll 1  &&  \hbox{(Josephson)}  \\
1 \ \ll \   E_C/E_J && \hbox{(Fock)}  \ .
\end{eqnarray*}
Recent experiments making use of strong interactions deep in the Josephson regime  have accomplished squeezing and macroscopic entanglement \cite{EsteveEtAl-squeezing}.  On the other hand, coherent tunneling dynamics and Bloch oscillations have been realized in completely non-interacting Bose gases \cite{GustavssonEtAl, FattoriEtAl}. The intermediate regime of moderate interactions (the Josephson - Rabi boundary regime) is virtually unexplored in experiment. This regime, where $ N^2E_C \sim E_J$, is of particular interest, as it contains most of the stationary Josephson modes, such as $0$ and $\pi$ phase modes, and the onset of macroscopic quantum self trapping \cite{RaghavanEtAl}.  Here, number and phase fluctuations are sensitive to the ratio of $ N^2E_C $ to $ E_J$, both in the ground state \cite{JavanainenIvanov} and, as we shall see, in thermal equilibrium at higher temperatures.

When a gas of ultracold  bosons is released from a double-well potential trap and recombined in free expansion, interference fringes, analagous to those of Young's double-slit experiment, are observed in the atomic density.   This does not necessarily mean that the double-well system was prepared in a coherent state, for individual images or ``shots" will feature interference fringes {\it even if the gases are initally independent} \cite{AndrewsEtAl}.   To ascertain that the experimental procedure prepares the system in a coherent state, one needs to repeat the experiment many times and compare the results.   If the fringes always lie in the same position, one may infer that the state is coherent, and ascribe a definite value to the relative phase between the condensates in the two wells.   

The decohering effect of temperature is seen in the fluctuations, from one shot to another, of the location of the interference fringes.  These fluctuations reduce the visibility of the interference fringes when the density profiles are averaged.  The fringe contrast in the average density profile is called the ``coherence factor" and, for double-well systems in the Josephson regime, is found to be a certain function of $k_BT/E_J$ \cite{PitaevskiiStringari}.   This function is used to calibrate the ``thermometer" of noise thermometry  \cite{GatiEtAl-NoiseThermometry, GatiEtAl-PrimaryNoise}.

In this article we study canonical thermal equilibrium states of $N \gg 1$ bosons in double-well and multi-well potentials, focusing on regimes where  both $E_J/k_BT \ll N$ and $N^2E_C/k_BT \ll N$.   This includes the Rabi-Josephson boundary regime, provided the temperature is high enough that $E_J/k_BT \ll N$.    We find that the coherence factor is sensitive to the ratios $E_J/k_BT $ and $N^2E_C/k_BT$.   Our results are rigorous inasmuch as they are derived from a general theorem about canonical statistics of $M$-mode boson models \cite{Gottlieb}.

The regimes we consider are not normally attained in atom interferometry experiments.  For example, the noise thermometry experiments reported in  \cite{GatiEtAl-NoiseThermometry,GatiEtAl-PrimaryNoise} involved trapping a few thousand $^{87}\mathrm{Rb}$ atoms at $15$-$80$ nK.  The parameter $ E_J/k_BT $ ranged between about $0.02$ and $20$, but the parameter $N^2E_C/k_BT$ was large
because the experiments were performed deep within the Josephson regime.  
%
However,  it should be possible to engineer bosonic Josephson junctions in the Rabi-Josephson boundary regime by taking advantage of Feshbach resonances to reduce the interaction parameter $E_C$ \cite{GustavssonEtAl, FattoriEtAl}. 

This paper is organized as follows.  In Section~\ref{applications section} we state our main results and proposals concerning noise thermometry of $2$-site bosonic Josephson junctions.   In Section~\ref{M-site section} we state the central result of this article, Theorem~\ref{SpecificTheorem}, a general result concerning  canonical statistics of $M$-site Bose-Hubbard models.   We return to the particular case $M=2$ in Section~\ref{2-site section} and discuss the fluctuations of density observables such as interference fringes in time-of-flight matter wave interferometry.     We outline a proof of Theorem~\ref{SpecificTheorem} in the Appendix.

\section{Noise thermometry with bosonic Josephson Junctions}
\label{applications section}

Double-well systems may be modeled using a two-mode approximation \cite{Javanainen}. In a symmetric two-well potential, the ground state is degenerate when the wells are separated by an infinitely high barrier:  the  {\it gerade} and {\it ungerade} modes 
have the same energy.  If the barrier between the wells is finite, tunneling lifts the ground state degeneracy.  Provided the tunneling barrier is not {\it too} low, the energy splitting of these low energy modes is small compared to the energy difference between them and the higher excited states, and, at low enough temperatures, there are effectively only two modes in play.

Two-mode approximations have been derived using either a ``semiclassical" or ``second-quantized" approach.   The former approach attempts to constrain the semiclassical Gross-Pitaevskii dynamics to a two-dimensional subspace \cite{SmerziEtAl,RaghavanEtAl}.  The latter approach begins with the second quantized Hamiltonian, and attempts to restrict it to one involving only two modes \cite{MilburnEtAl}, which might be the lowest energy solutions of the Gross-Pitaevskii equation \cite{JavanainenIvanov}, or which might be found by self-consistent variational minimization of the energy over all suitable two mode approximations \cite{SpekkensSipe}.     The relationship between the second-quantized and the semiclassical theories is discussed in \cite{AnglinDrummondSmerzi}. 

Two-mode models become quite sophisticated \cite{AnanikianBergeman}.  The simplest one is the  2-site Bose-Hubbard model, whose Hamiltonian is 
\begin{equation}
\label{Bose-Hubbard 2-site}
H_N \equals 
- E_J \hx \ + \  (N^2 E_C/4)  \ \hz^2 \ .
\end{equation}
In this formula, the operators 
$
\hx = \tfrac{1}{N}(a_1^{\dagger}a_2 + a_2^{\dagger}a_1)$ and $\hz= \tfrac{1}{N}(a_1^{\dagger}a_1 -  a_2^{\dagger}a_2)
$ are understood to operate on the $N$-particle component of the boson Fock space.
The observable $\hz$ is the relative number imbalance between the wells.  The observable $\hx$ is the relative occupation difference of the {\it gerade} and {\it ungerade} modes, which is related to relative phase. 

Canonical thermal equilibrium states of the $2$-site Bose-Hubbard Hamiltonian can be approximated by certain mixtures of coherent states.    
We shall show that this approximation is rigorously justified for regimes where $E_J/k_BT \ll N$ and $N^2E_C/k_BT \ll N$ (and $N$ is large).  
Define the dimensionless parameters
\begin{eqnarray*}
 \delta  & = & E_J/k_BT \\
   \varepsilon & = &  N^2E_C/4k_BT \ .
\end{eqnarray*}
For regimes where $\delta, \varepsilon \ll N$ , we will derive the following formulas for the coherence factor (\ref{formula for coherence factor}) and the second moment of $\hz$ (\ref{formula for z^2}). 

The coherence factor $\alpha$ is defined to be the fringe contrast in the ensemble averaged density profile of a double-well interference experiment \cite{PitaevskiiStringari, GatiEtAl-PrimaryNoise, GatiOberthaler}. 
We will show that 
\begin{equation}
\alpha
\equals
\frac{ \int_{-1}^1 x  I_0 \big( \varepsilon(1-x^2)/4 \big)e^{\delta x + \varepsilon x^2/4} dx }
{\int_{-1}^1  I_0 \big( \varepsilon(1-x^2)/4 \big)e^{\delta x+  \varepsilon x^2/4} dx } \ ,
\label{formula for coherence factor} 
\end{equation}
where $I_0$ denotes the modified Bessel function of the first kind (of order zero).  
In the non-interacting case, when $\varepsilon = 0$, formula (\ref{formula for coherence factor}) reduces to   
\begin{equation}
\label{noninteracting formula}
\alpha_{\delta, 0}
\equals 
\coth(\delta)-1/\delta\  .
\end{equation}
In the strongly repulsive case, when $\varepsilon \gg 1$, the term $I_0 \big( \varepsilon(1-x^2)/4 \big)e^{\varepsilon x^2/4 }$    is nearly proportional to $1/\sqrt{1-x^2}$ over much of the domain of integration, and formula (\ref{formula for coherence factor}) tells us that 
\begin{equation}
\label{pitaevskii-stringari formula}
\alpha_{\delta, \infty} 
\ \approx\ 
\frac{ \int_{-1}^1 x \frac{e^{\delta x}}{\sqrt{1-x^2}} dx }{\int_{-1}^1  \frac{e^{\delta x}}{\sqrt{1-x^2}} dx } \equals \frac{I_1(\delta)}{I_0(\delta)}\  .
\end{equation}
This agrees with the semiclassical formula for the coherence factor in the Josephson regime \cite{PitaevskiiStringari,GatiEtAl-NoiseThermometry,GatiEtAl-PrimaryNoise}.  

In a symmetric double-well potential, the expected value  of the population imbalance $\hat{z}$ is zero, i.e., $\langle \hat{z} \rangle=0$. 
We will show that the variance of $\hz$ is 
\begin{equation}
\langle \hz^2 \rangle
\equals
\frac{ \int_0^1 z^2 I_0 \big( \delta \sqrt{1-z^2}\big) e^{-\varepsilon z^2/2 }dz }{\int_0^1  I_0 \big( \delta\sqrt{1-z^2}\big) e^{-\varepsilon z^2/2 }dz} \ .
\label{formula for z^2}
\end{equation}

In Figures 1 and 2,  $\alpha$ and $ \langle \hz^2 \rangle$ are plotted against $k_BT/E_J$ on a logarithmic scale, for various values of 
\[
    \Lambda \equals \varepsilon/\delta \equals N^2 E_C /4 E_J\ .
\]

\begin{figure}
\includegraphics[angle=0,width=0.9\columnwidth]{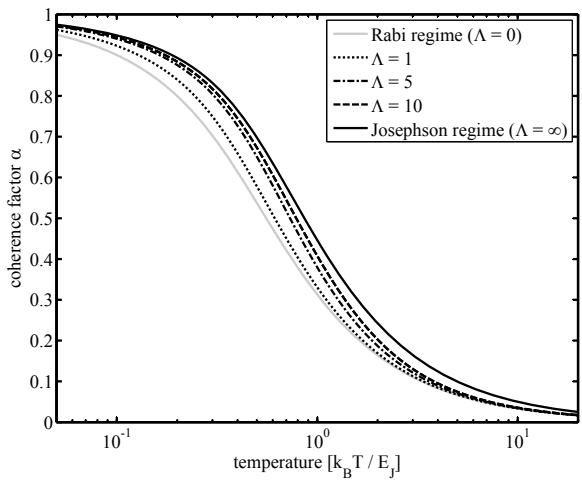}
\caption{Coherence factor $\alpha $ vs. temperature for several values of $\Lambda= N^2 E_C /4 E_J$ (logarithmic scale).}
\label{Fig:coherence_factor}
\end{figure}

\begin{figure}
\includegraphics[angle=0,width=0.9\columnwidth]{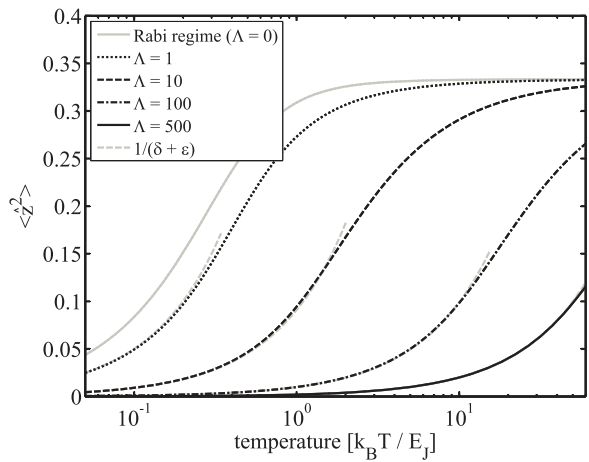}
\caption{Variance of the normalized population imbalance $\langle \hat{z}^2 \rangle$ vs. temperature  for various values of $\Lambda= N^2 E_C /4 E_J$.  The gray dashed lines indicate the ``low temperature" approximation (\ref{EQ:low_temp_approx}).}
\label{FIG:populations}
\end{figure}

The method of ``noise thermometry" developed in \cite{GatiEtAl-NoiseThermometry, GatiEtAl-PrimaryNoise} uses statistical estimates of $\alpha$, obtained by replicating a double-well experiment under identical conditions, to determine $k_BT/E_J$.   If $E_J$ is known, or estimable, the temperature $T$ can be deduced, even when this temperature is so low that it cannot be found by the usual technique (fitting a gaussian to the ``wings" of the density profile after some time-of-flight).  This method is suitable for the Josephson regime $E_J \ll N^2E_C$, where the parameter $\Lambda$ is formally equal to $\infty$. 

To perform noise thermometry in the Rabi-Josephson boundary regime, where $0 \ll \Lambda \ll \infty$, one needs to know both $k_BT/E_J$ and $\Lambda$.   
Estimates of $k_BT/E_J$ and $\Lambda$ can be deduced from statistical estimates of the coherence factor $\alpha$ and the variance $\langle \hat{z}^2 \rangle$ of the number fluctuations, thanks to formulas (\ref{formula for coherence factor}) and (\ref{formula for z^2}).   

There are a couple of benefits of doing noise thermometry in the Rabi-Josephson boundary regime:

\noindent 1.\quad When a bosonic Josephson junction is fashioned in the laboratory, one usually has more accurate knowledge of the parameter $E_C$ than the parameter $E_J$ (the tunneling energy $E_J$ is quite difficult to estimate, due to its exponential sensitivity to the precise geometry of the double-well potential).    By performing noise thermometry in the Rabi-Josephson boundary regime, one can take advantage of one's knowledge of $E_C$ to estimate  $E_J$ as well as $T$. 

\noindent 2.\quad  If one does happen to know $E_J$ with some accuracy,
 one obtains {\it two} estimates of the temperature.  If these estimates differ significantly, it may be evidence that the experimental procedure has failed to prepare the double-well system in a canonical thermal equilibrium state.  The assumption that replication of the experiment is  sampling from the canonical ensemble ought to be tested, because some ways of preparing the system can fix it in a non-canonical equilbrium state, for example, if the double-well is formed by ramping up a potential barrier too quickly \cite{GatiOberthaler}.


\section{Canonical statistics of the M-site Bose-Hubbard model}  
\label{M-site section}

The $M$-site Bose-Hubbard Hamiltonian for bosons with nearest-neighbor hopping is 
\[
- \ J \sum_{i=1}^{M-1} \big( a_i^{\dagger}a_{i+1}+a_{i+1}^{\dagger}a_i \big) \ + \  \frac{U}{2}\sum_{i=1}^M a_i^{\dagger}a_i^{\dagger}a_i a_i \ .
\] 
We are going to discuss systems of exactly $N$ bosons, and take a limit $N \longrightarrow \infty$.  Accordingly, we will consider the restriction of the above Hamiltonian to the $N$-particle subspaces of the boson Fock space, and allow the parameters $J$ and $U$ to depend on $N$.  
Let $\PP_N$ denote the orthogonal projector onto the $N$-particle component of the boson Fock space over $\CCC^M$, and let  
\begin{equation}
\label{Bose-Hubbard-M}
\Big( - J_N \sum_{i=1}^{M-1} \big( a_i^{\dagger}a_{i+1}+a_{i+1}^{\dagger}a_i \big) \ + \   \frac{U_N}{2}\sum_{i=1}^M a_i^{\dagger}a_i^{\dagger}a_i a_i \Big) \PP_N
\end{equation}
be the the $N$-boson Hamiltonian $H_N$.  

The density operator 
\[
e^{- H_N/k_BT} / \tr \big(e^{- H_N/k_BT}\big) 
\]
represents the canonical ensemble of $N$ bosons in thermal equilibrium at temperature $T$, in the sense that 
\begin{equation}
\label{the ensemble average}
  \langle X \rangle_{N,T} \ =\   \tr \big( X  e^{- H_N/k_BT}  \big) / \tr \big(e^{- H_N/k_BT}\big) 
\end{equation}
is the expected value of any observable $X$ when the system is in the canonical thermal equilibrium state.
We are going to show that this state may be approximated by a mixture of coherent states, provided that $N \gg 1$ and both $NJ_N/k_BT,\ N^2U_N/k_BT \ll N$.

We parameterize the pure states of the $M$-site system by the product of the standard $(M-1)$-dimensional simplex and the $M$-dimensional torus.  Let $\Delta_M$ 
denote the standard $(M-1)$-dimensional simplex
\[
\Big\{  p = (p_1,p_2,\ldots,p_M) \in \RRR^M | \  p_1+\cdots + p_M = 1\ , \ \ p_i \ge 0 \Big\}
\]
and let $[0,2\pi)^{M}$ denote 
\[
\Big\{ \phi =  (\phi_1,\phi_2,\ldots,\phi_M) | \ \phi_i \in [0,2\pi)\  \hbox{for } i = 1,2,\ldots,M  \Big\}.
\]
To each point $(p,\phi) \in  \Delta_M \times [0,2\pi)^M$ we associate the unit vector
\begin{equation}
\label{psi}
\psi_{p,\phi}
\equals 
\big(
  \sqrt{p_1}e^{i\phi_1}, \sqrt{p_2}e^{i\phi_2},\ldots,  \sqrt{p_M}e^{i\phi_M}
\big)
 \ .
\end{equation}
The parameterization $(p,\phi) \mapsto \psi_{p,\phi}$ is many-one, because a global change of phase in $\phi$ does not change $\psi_{p,\phi}$.

Let $\mu_M(dp)$ denote the uniform probability measure on $\Delta_M$; in particular, $\mu_2$ is equivalent to the length measure $dp$ on the unit interval $p\in [0,1]$.   Let $\nu_M(d\phi)$ denote the uniform probability measure $(2\pi)^{-M} d\phi_1 d\phi_2 \cdots d\phi_M$ on $[0,2\pi)^M$.

\begin{theorem}
\label{SpecificTheorem}
Let $J_N$ and $U_N$ be two sequences of parameter values.  For each $N$, let $H_N$ denote the operator (\ref{Bose-Hubbard-M}), and let $\langle X \rangle_{N,T}$ denote the ensemble average (\ref{the ensemble average}) for the canonical ensemble at temperature $T$.

If
\begin{equation}
\label{the limits}
 \lim_{N \rightarrow \infty} \frac{NJ_N}{k_BT} \equals  \delta \quad \mathrm{and} \quad 
 \lim_{N \rightarrow \infty} \frac{N^2U_N}{4k_BT} \equals   \varepsilon \ ,
\end{equation}
then, for any vectors $\chi_1,\ldots \chi_k,\chi'_1,\ldots, \chi'_k \in \CCC^M$, 
\begin{widetext}
\begin{eqnarray}
& \lim\limits_{N \rightarrow \infty} &
\frac{1}{N^k}
\big\langle a^{\dagger}_{\chi_1}a^{\dagger}_{\chi_2}\cdots a^{\dagger}_{\chi_k} a_{\chi'_1}a_{\chi'_2}\cdots a_{\chi'_k} \big\rangle_{N,T} 
\nonumber \\
& = & 
   \int_{\Delta_M}
\int_{ [0,2\pi)^M }
\prod_{i=1}^k  \langle \chi_i |  \psi_{p,\phi} \rangle  \langle \psi_{p,\phi} | \chi'_i \rangle 
\ \frac{ 
        \exp\Big[ 2 \delta \sum\limits_{i=1}^{M-1} \sqrt{p_i p_{i+1} } \cos(\phi_{i+1}-\phi_i) -  \varepsilon \sum\limits_{i=1}^{M} p_i^2 \Big]
   }{
      \int\int   e^{  2 \delta \sum \sqrt{p'_i p'_{i+1} }  \cos(\phi'_{i+1}-\phi'_i)  - \varepsilon  \sum {p'_i}^2} 
       \mu_M(dp')  \nu_M(d\phi'_i) 
   } 
    \ \nu_M(d\phi)\mu_M (dp)\ .
    \nonumber \\
\label{The Limit (M-site)}
\end{eqnarray}
\end{widetext}

\end{theorem}  

This theorem can be deduced from propositions concerning Finetti representations for canonical states of $M$-mode bosons \cite{Gottlieb}.  A proof is outlined in the appendix.


\section{Derivation of formulas  (\ref{formula for coherence factor}) and (\ref{formula for z^2})}
\label{2-site section}

%
The $2$-site Bose-Hubbard Hamiltonian is \cite{AnglinDrummondSmerzi}
\[
-  \frac{E_J}{N} \big(  a_1^{\dagger}a_2+a_2^{\dagger}a_1 \big) \PP_N \ + \  \frac{E_C}{4} \big( a_1^{\dagger}a_1^{\dagger}a_1a_1 + a_2^{\dagger}a_2^{\dagger}a_2a_2 \big) \PP_N
\]
 ($\PP_N$ restricts the operators to the $N$-particle component of the Fock space).
%
%
When expressed in terms of the observables
\begin{eqnarray*}
\hx & = &  \tfrac{1}{N}(a_1^{\dagger}a_2 + a_2^{\dagger}a_1)\PP_N
\\
\hz & = &  \tfrac{1}{N}(a_1^{\dagger}a_1 -  a_2^{\dagger}a_2)\PP_N
\end{eqnarray*}
on the $N$-boson space, this Hamiltonian  only differs by 
the constant $E_CN ( N-2 )/4$ from the Hamiltonian   
(\ref{Bose-Hubbard 2-site}).

We are going to rewrite formula (\ref{The Limit (M-site)}) for $M=2$, the double-well case.  
Changing variables 
\[   p_1 = \tfrac12 + \tfrac12 z, \ p_2 = \tfrac12 - \tfrac12 z, \ \phi = \phi_2 - \phi_1,\ \phi' = \phi_1 + \phi_2 \] 
in  formula (\ref{psi}), we write 
\[
 \psi_{p,\phi} \equals e^{i\phi'/2}  \big( \tfrac{\sqrt{1+z}}{\sqrt{2}}e^{-i\phi/2}, \  \tfrac{\sqrt{1-z}}{\sqrt{2}} e^{i\phi/2} \big)  \ .
\]
Define 
\begin{equation}
\label{vec-u}
\vec{u}_{z,\phi} \equals     
\big( \tfrac{\sqrt{1+z}}{\sqrt{2}}e^{-i\phi/2}, \  \tfrac{\sqrt{1-z}}{\sqrt{2}} e^{i\phi/2} \big)    
\end{equation}
for $(z,\phi) \in [-1,1] \times [0,2\pi)$.  The operators $a^{\dagger}_1$ and $a^{\dagger}_2$ in $H_N$ are identified with the creation operators for the vectors $\vec{u}_{1,0}=(1,0)$ and $\vec{u}_{-1,0}=(0,1)$, respectively.
Let us also define the probability density functions 
\begin{equation}
\label{probability density}
 P_{\delta,\varepsilon}(z,\phi) \equals \frac{ \exp\big( \delta  \sqrt{1-z^2 }  \cos\phi  - \varepsilon  z^2 / 2 \big) }{ \int \int e^{ \delta  \sqrt{1-{z'}^2 }  \cos\phi'  - \varepsilon  {z'}^2/2} d\phi' dz'  } 
\end{equation}
on $ [-1,1] \times [0,2\pi)$.  
Changing variables in formula (\ref{The Limit (M-site)}) 
we find that 
\begin{eqnarray}
& \lim\limits_{N \rightarrow \infty} &
\frac{1}{N^k}
\big\langle a^{\dagger}_{\chi_1}\cdots a^{\dagger}_{\chi_k} a_{\chi'_1}\cdots a_{\chi'_k} \big\rangle_{N,T} 
\nonumber \\
&=&  \int_0^{2\pi}  \int_{-1}^1\prod_{i=1}^k  \langle \chi_i |  \vec{u}_{z,\phi} \rangle \langle \vec{u}_{z,\phi} | \chi'_i \rangle P_{\delta,\varepsilon}(z,\phi) dz  d\phi 
\nonumber \\
\label{The Limit (2-site)}
\end{eqnarray}
in the limit $ N \longrightarrow \infty$ with
\begin{equation}
\label{way to the limit}
  E_J/k_BT \longrightarrow \delta,\quad N^2E_C/4k_BT \longrightarrow \varepsilon\ .
\end{equation}

%
%
\subsection{Population imbalance}

Formula (\ref{The Limit (2-site)}) may be applied directly to the observable $\hz$.  In a symmetric double-well, $\langle \hz \rangle=0$.  Higher moments of $\hz$ are those of 
the probability distribution
\[
\frac{ 
      I_0 \big( \delta \sqrt{1-z^2}\big) e^{-\varepsilon z^2/2}
      }
      {
      \int_0^1  I_0 \big( \delta\sqrt{1-(z')^2}\big) e^{-\varepsilon (z')^2/2}dz'
      }
      \ dz \ ,
\]
that is, in the limit (\ref{way to the limit})  for each fixed $k$, 
\[
\lim\  \langle \hz^k \rangle_{N,T}
\equals 
\frac{ \int_{-1}^1 z^k   I_0 \big( \delta \sqrt{1-z^2}\big) e^{-\varepsilon z^2/2} dz }
{\int_{-1}^1  I_0 \big( \delta \sqrt{1-z^2}\big) e^{-\varepsilon z^2/2} dz}  \ .
\]
We demonstrate this for $k=2$:
\begin{eqnarray*}
\langle \hz^2 \rangle_{\delta,\varepsilon} & := &  \lim\  \langle \hz^2 \rangle_{N,T} \\
& = & 
\lim\  \frac{1}{N^2} \big\langle 
(a_1^{\dagger}a_1 -  a_2^{\dagger}a_2)^2 
\big\rangle_{N,T}    \nonumber \\
& = & 
 \lim\  \frac{1}{N^2} \big\langle 
a_1^{\dagger 2} a_1^2 + a_2^{\dagger 2} a_2^2 - 2 a_1^{\dagger} a_2^{\dagger}  a_1  a_2
\big\rangle_{N,T}    \nonumber \\
 &  & 
\qquad + \ 
 \lim\  \frac{1}{N^2} \big\langle 
a_1^{\dagger}a_1 +  a_2^{\dagger}a_2
\big\rangle_{N,T}    \nonumber \\
& =  &  
\int_0^{2\pi}   \int_{-1}^1
\Big( \big| \langle \vec{u}_{1,0} |  \vec{u}_{z,\phi} \rangle \big|^2 - \big| \langle \vec{u}_{-1,0} |  \vec{u}_{z,\phi} \rangle \big|^2\Big)^2  
\nonumber \\
&  & 
\qquad \times\  P_{\delta,\varepsilon}(z,\phi)\ dz d\phi   \nonumber \\
& =  & 
\int_0^{2\pi}   \int_{-1}^1  z^2    P_{\delta,\varepsilon}(z,\phi)\ dz d\phi
\\
& = & 
\frac{ \int_{-1}^1 z^2   I_0 \big( \delta \sqrt{1-z^2}\big) e^{-\varepsilon z^2/2} dz }
{\int_{-1}^1  I_0 \big( \delta \sqrt{1-z^2}\big) e^{-\varepsilon z^2/2} dz}    \ .
\end{eqnarray*}
This proves formula (\ref{formula for z^2}) for $ \langle \hz^2 \rangle_{\delta,\varepsilon} $.
Figure~\ref{FIG:populations} shows that 
\begin{equation}   
\langle \hz^2 \rangle_{\delta,\varepsilon} \approx \frac{1}{\delta + \varepsilon}
\label{EQ:low_temp_approx}
\end{equation}
is a good approximation at lower temperatures.

%
%
\subsection{Coherence factor}

In a time-of-flight (TOF) experiment on double-wells, the potential trap is suddenly shut off and the gas expands into free space for awhile before it is imaged.  
The images constitute a measurement of the ``integrated density" observable  $\int a^{\dagger}(\mathbf{r}) a(\mathbf{r}) dr_3$, where $a^{\dagger}(\bf{r})$ denotes the usual field operator at $\mathbf{r}=(r_1,r_2,r_3)$, and the integral is over the spatial coordinate $r_3$ parallel to the imaging light beam and perpendicular to the line that passes through the two wells, the $r_1$-axis.  
We turn our attention to the density observables $a^{\dagger}(\mathbf{r}) a(\mathbf{r})$ and their linear combinations.

Moments of such density observables are easily computed if atom-atom interactions during the TOF are neglected.  
In a two-mode approximation, each vector $\vec{u}_{z,\phi} \in \CCC^2$ is identified with some wavefunction $\Psi_{z,\phi}(\mathbf{r})$. 
 The vectors $\vec{u}_{-1,0}$ and $\vec{u}_{1,0}$ are identified with the wavefunctions of the ``left" and ``right" wells, respectively.  The specific map $\vec{u} \mapsto \Psi$ depends on the two-mode approximation adopted, but the precise form of the initial left and right well wavefunctions hardly affects the interference pattern observed after a long TOF, and we may simply assume that $\Psi_{-1,0}$ and $\Psi_{+1,0}$ are gaussian wave packets centered at $(-d/2,0,0)$ and $(d/2,0,0)$ \cite{ImambekovGritsevDemler}.
     After a long enough \footnote{ $t$ so large that $\sqrt{\hbar t/m}$ is much greater than the width of the wells.} time $t$ of free expansion, 
the wavefunction $ \Psi_{\pm 1,0}(\mathbf{r},t) $, which describes the state of an atom that was initially in right (+) or left (-) well, will be nearly proportional to 
\begin{equation}
\label{quite like this}
  \exp\Big( -i \frac{m}{2\hbar t} \big(\|\mathbf{r}\|^2 + d^2/4 \big) \Big) \exp\Big(  \mp i  \frac{md}{2\hbar t} r_1 \Big)
\end{equation}
over the region where  the density is imaged.  
Let $\Psi_{z,\phi}(\mathbf{r},t)$ denote 
\[
  \tfrac{\sqrt{1+z}}{\sqrt{2}}e^{-i\phi/2}\Psi_{-1,0} (\mathbf{r},t) \ + \    \tfrac{\sqrt{1- z}}{\sqrt{2}}e^{i\phi/2} \Psi_{+ 1,0}(\mathbf{r},t) \ .
\]
Supposing that atom-atom interactions during the period of expansion may be neglected, the state of the many-boson system at time $t$ is just the one freely induced by the $1$-particle map $\vec{u}_{z,\phi} \mapsto  \Psi_{z,\phi}$, and Theorem~\ref{SpecificTheorem} implies that 
\begin{eqnarray}
& \lim\limits_{N \rightarrow \infty} &
\frac{1}{N^k}
\big\langle a^{\dagger}(\mathbf{r}_1)\cdots a^{\dagger} (\mathbf{r}_k) a (\mathbf{r'}_1) \cdots a (\mathbf{r'}_k)  \big\rangle_{N,T} \nonumber \\
&= &  \int_0^{2\pi}  \int_{-1}^1   \prod_{i=1}^k   \Psi_{z,\phi}(\mathbf{r}_i,t) \overline{\Psi_{z,\phi}}(\mathbf{r'}_i,t) P_{\delta,\varepsilon}(z,\phi)
dz  d\phi 
\nonumber \\
\label{The Limit in Space}
\end{eqnarray}
for all $k$ and points $\mathbf{r}_1,\ldots,\mathbf{r}_k$ and $\mathbf{r'}_1,\ldots,\mathbf{r'}_k$,
in the limit (\ref{way to the limit}).  
Substituting (\ref{quite like this}) into (\ref{The Limit in Space}) and proceeding formally, one finds that 
\begin{eqnarray}
& \lim\limits_{N \rightarrow \infty} &
\frac{1}{N}
\big\langle a^{\dagger} (\mathbf{r}) a (\mathbf{r})  \big\rangle_{N,T}   \nonumber \\
& =  &  \int_0^{2\pi}  \int_{-1}^1 
\big| \Psi_{z,\phi}(\mathbf{r},t)\big|^2  \ P_{\delta,\varepsilon}(z,\phi) dz  d\phi  \nonumber \\
& \propto & 1 \ + \ 
\int_{-1}^1  \int_0^{2\pi} 
\sqrt{1-z^2} \cos\Big( \frac{md}{\hbar t} r_1 - \phi \Big)P_{\delta,\varepsilon}(z,\phi)  d\phi dz  \nonumber \\
& = &  1 \ + \ \alpha_{\delta,\varepsilon} \cos\Big( \frac{md}{\hbar t} r_1\Big) \ ,
\label{Limit of the density}
\end{eqnarray}
where 
\begin{equation}
\label{alpha}
 \alpha_{\delta,\varepsilon}  \equals \int_{-1}^1  \int_0^{2\pi}   
\sqrt{1-z^2} \cos \phi\ P_{\delta,\varepsilon}(z,\phi)  d\phi dz  \ .
\end{equation}
Finally, fomula (\ref{formula for coherence factor}) for $\alpha_{\delta,\varepsilon}$ is obtained by changing variables $x=\sqrt{1-z^2}  \cos\phi,\ y=\sqrt{1-z^2}  \sin\phi$ and integrating over $y$.

 Formula (\ref{Limit of the density}) shows that the interference pattern will feature fringes with spacing $ht/md$ and contrast equal to the coherence factor $\alpha_{\delta,\varepsilon}$.
In particular, formula (\ref{Limit of the density})  implies that 
\[
\lim_{N \rightarrow \infty}\frac{1}{N}
\Big\langle \int a^{\dagger} (\mathbf{r}) a (\mathbf{r}) e^{i \mathbf{k}\cdot \mathbf{r} } d\mathbf{r}   \Big\rangle_{N,T}  
\]
is proportional to  $ \alpha_{\delta,\varepsilon} $ when $\mathbf{k} = (md/\hbar t ,0,0)$.  Thus the coherence factor can be estimated by averaging, over many replicates of a TOF experiment,  the Fourier coefficient of the imaged density profiles at wave-vector $\mathbf{k}$.


\bigskip

\section{Conclusion}

We have studied the canonical statistics of phase and number in the $M$-site Bose-Hubbard model (\ref{Bose-Hubbard-M}).  
Theorem~\ref{SpecificTheorem} provides a 
convenient way to approximate canonical thermal equilibrium states by mixtures of coherent states.   
From Theorem~\ref{SpecificTheorem}  we have derived formulas (\ref{formula for coherence factor})  and  (\ref{formula for z^2})   for the coherence factor $\alpha$ and the variance of the relative population imbalance in symmetric double-well bosonic Josephson junctions. 
These formulas are valid in the Rabi-Josephson boundary regime, provided $N \gg 1$ and $E_J/k_BT \ll N$.

We have proposed a way to perform noise thermometry in the Rabi-Josephson boundary regime.  In this regime, canonical statistics depend on two parameters, e.g., the dimensionless parameters $E_J/k_BT$ and $\Lambda = N^2E_C/4E_J$.   Statistical estimates of the coherence factor and the variance of the number fluctuations can be used to  obtain empirical estimates of $E_J$ and $T$, and to 
test the assumption that the system is being prepared in a canonical equilibrium state.

\bigskip

\noindent {\bf Acknowledgment:}\quad 
A. G. is supported by the Vienna Science and Technology Fund project ``Correlation in Quantum Systems".    This work was done under the auspices of Joerg Schmiedmayer's Atom Chip Lab.   We thank Igor Mazets for helpful comments.


\bibliographystyle{apsrev}

\section{Appendix: proof of Theorem~\ref{SpecificTheorem}}

Recall that $\PP_N$ denotes the projector whose range is the $N$-particle component of the boson Fock space 
over $\CCC^M$, and recall the notation introduced around formula (\ref{psi}).
Proposition~2 of \cite{Gottlieb} implies that
\begin{eqnarray}
& \lim\limits_{N \rightarrow \infty} &
\frac{1}{N^k} \tr\big( ( a^{\dagger}_{\chi_1}\cdots a^{\dagger}_{\chi_k} a_{\chi'_1}\cdots a_{\chi'_k} ) \PP_N \big) / \tr(\PP_N) 
\nonumber \\
& = & 
\int_{\Delta_M} \int_{ [0,2\pi)^M }    
\prod_{i=1}^k  \langle \chi_i |  \psi_{p,\phi} \rangle  \langle \psi_{p,\phi} | \chi'_i \rangle 
    \ \nu_M(d\phi)\mu_M (dp)
    \nonumber \\
\label{howdy}
\end{eqnarray}
for any vectors $\chi_1,\ldots \chi_k,\chi'_1,\ldots, \chi'_k \in \CCC^M$.
Indeed, formula (\ref{howdy}) holds even if the product of the $a^{\dagger}_{\chi_i}$ and $a_{\chi'}$ is not normally ordered.  

Writing the operator defined in (\ref{Bose-Hubbard-M}) as 
\[
  H_N \equals \big( - J_N \hat{X}_1 +  (U_N/2)   \hat{X}_2 \big)\PP_N \ ,
\]
we may write 
\begin{widetext}
\begin{eqnarray*}
e^{- H_N/k_BT} 
& = &
\sum_{n_1 = 0}^{\infty} \sum_{n_2=0}^{\infty} \frac{(-1)^{n_2}}{(n_1+n_2)!}\Big(\frac{N J_N}{k_BT}\Big)^{n_1} \Big(\frac{N^2 U_N}{2 k_BT}\Big)^{n_2} \frac{1}{N^{n_1+2n_2}} 
 ( \hat{X}_1^{n_1} \hat{X}_2^{n_2} +  \hat{X}_1^{n_1-1} \hat{X}_2 \hat{X}_1 \hat{X}_2^{n_2-1 }  + \cdots + \hat{X}_2^{n_2}\hat{X}_1^{n_1} ) \PP_N\ .
\end{eqnarray*}
\end{widetext}
We are going to take a limit of the trace of both sides of the preceding equation.
Formula (\ref{howdy}) implies that a limit such as 
\[
    \lim_{N \rightarrow \infty}  \frac{1}{N^{n_1+2n_2} } \tr \big(  \hat{X}_1^{n_1-1} \hat{X}_2 \hat{X}_1 \hat{X}_2^{n_2-1 } \PP_N  \big) / \tr(\PP_N)
\]
is equal to the same limit for the normally ordered form of the operator, i.e., the limit here is identical to 
\begin{equation}
\label{lim1}
    \lim_{N \rightarrow \infty}  \frac{1}{N^{n_1+2n_2} } \tr \big( \colon \hat{X}_1^{n_1} \hat{X}_2^{n_2}  \colon \PP_N  \big) / \tr(\PP_N)\ .
\end{equation}
According to formula (\ref{howdy}), the limit in (\ref{lim1}) equals 
\[
\int_{\Delta_M} \int_{ [0,2\pi)^M }    
  f(p,\phi)^{n_1} g(p,\phi)^{n_2} \ \nu_M(d\phi)\mu_M (dp)\ ,
\]
where $     f(p,\phi) =  2 \sum\limits_{i=1}^{M-1} \sqrt{p_ip_{i+1}}  \cos(\phi_{i+1} - \phi_i) $  
and
  $ g(p,\phi) =  \sum\limits_{i=1}^M p_i^2 $.      
Therefore, 
\begin{widetext}
\begin{eqnarray*}
\lim_{N \rightarrow \infty} \frac{ \tr \big( e^{- H_N/k_BT} \big) }{\tr(\PP_N)}
& = &
\sum_{n_1 = 0}^{\infty} \sum_{n_2=0}^{\infty} \frac{(-1)^{n_2}}{(n_1+n_2)!}\delta^{n_1} \varepsilon^{n_2} 
\binom{ n_1+n_2 }{n_1} \int_{\Delta_M} \int_{ [0,2\pi)^M }    
  f(p,\phi)^{n_1} g(p,\phi)^{n_2} \ \nu_M(d\phi)\mu_M (dp)
\\
& = &
\int_{\Delta_M} \int_{ [0,2\pi)^M }   
\sum_{n_1 = 0}^{\infty} \sum_{n_2=0}^{\infty} \frac{1}{n_1!}\frac{1}{n_2!}
   (\delta f(p,\phi))^{n_1} (-\varepsilon g(p,\phi)) ^{n_2} 
 \ \nu_M(d\phi)\mu_M (dp)
  \\
& = &
\int_{\Delta_M} \int_{ [0,2\pi)^M }   
e^{  \delta f(p,\phi) - \varepsilon g(p,\phi)} 
 \ \nu_M(d\phi)\mu_M (dp)\ .
\end{eqnarray*}
Similarly, 
\[
  \lim_{N \rightarrow \infty} \frac{ \tr \big( (a^{\dagger}_{\chi_1}\cdots a^{\dagger}_{\chi_k} a_{\chi'_1}\cdots a_{\chi'_k} ) e^{- H_N/k_BT} \big) }{\tr(\PP_N)} \equals  \int_{\Delta_M} \int_{ [0,2\pi)^M }   
\prod_{i=1}^k  \langle \chi_i |  \psi_{p,\phi} \rangle  \langle \psi_{p,\phi} | \chi'_i \rangle e^{  \delta f(p,\phi) - \varepsilon g(p,\phi)} 
 \ \nu_M(d\phi)\mu_M (dp)\ .
\]
\end{widetext}
The last two equations imply formula  (\ref{The Limit (M-site)}).

\end{document}